\documentclass[final,5p,times,twocolumn]{elsarticle}

\usepackage{lipsum}

\usepackage{lineno}

\usepackage{amssymb}
\usepackage{amsmath}
\usepackage{here}
\usepackage{braket}
\usepackage{bm}
\usepackage[driverfallback=dvipdfmx,colorlinks=true, bookmarks=true,bookmarksnumbered=true,bookmarkstype=toc]{hyperref}
\usepackage{braket}

\usepackage[dvipsnames]{xcolor}

\journal{}

\newcounter{daggerfootnote}

\begin{document}

\begin{frontmatter}



\title{An accelerator experiment for junior
and senior high school students to improve students’ involvement in fundamental physics}
\author[tohoku]{K.S. Tanaka\corref{Tanaka}}
\cortext[Tanaka]{present address:  Paul Scherrer Institute (PSI), Switzerland.}
\ead{tanaka@kaduo.jp}
\author[tokyotech]{K. Harada}
\author[riken]{T. Hayamizu}
\author[tohokusci]{R. Kita}
\author[tohokusci]{R. Kono}
\author[tohokusci]{K. Maruta}
\author[cns]{H. Nagahama}
\author[utokyo]{N. Ozawa}
\author[cns]{Y. Sakemi}
\author[tohokusci]{R. Sugimori\fnref{sugimori}}
\fntext[sugimori]{Deceased 23 January 2020}

\address[tohoku]{Cyclotron and Radioisotope Center (CYRIC), Tohoku University, 6-3 Aramaki-aza Aoba, Aoba-ku, Miyagi 980-8578, Japan}
\address[tokyotech]{Department of Physics, Tokyo Institute of Technology, 2-12-1 Ookayama, Meguro-ku, Tokyo 152-8551, Japan}
\address[riken]{Nishina Center for Accelerator-Based Science, RIKEN, 2-1 Hirosawa, Wako, Saitama 351-0198, Japan}
\address[tohokusci]{Faculty of Science, Tohoku University, 6-3 Aramaki-aza Aoba, Aoba-ku, Miyagi 980-8578, Japan}
\address[cns]{Center for Nuclear Study, Graduate School of Science, The University of Tokyo, 2-1 Hirosawa, Wako, Saitama 351-0198, Japan}
\address[utokyo]{Graduate School of Science, The University of Tokyo, 7-3-1, Hongo, Bunkyo-ku, Tokyo 113-0033, Japan}

\renewcommand{\thefootnote}{\fnsymbol{footnote}}
\title{An accelerator experiment for junior
and senior high school students to improve students’ involvement in fundamental physics}

\begin{abstract}
  In Japan, research activities by junior and senior high school students show an upward trend. However, there are limited examples of research activities in the field of elementary particles and atoms. This is due to the difficulty associated with procuring research tools such as accelerators or particle detectors.
  Therefore, we hosted the ``Accel Kitchen'' in 2018 and 2019 at Cyclotron and Radioisotope Center (CYRIC) in Tohoku University where junior and senior high school students could participate in ongoing research of particle and atomic physics.
  At each workshop, 12 junior and senior high school students participated in the beam experiment, including the production of francium atoms (Fr) by the fusion reaction of oxygen and gold, optimizing the transport of the ion beam and identifying the alpha decay nuclei, and laser trapping of Fr for two days. Each group that was involved in the experiment was supported by researchers and university students who acted as mentors. 
  This was the first opportunity for junior and senior high school students to know about the particle beam experiment in Japan.
\end{abstract}  

 
 \end{frontmatter}

\section{\label{sec:introduction}Introduction}
In recent years, students’ interest in science at school has witnessed a decline in Japan~\cite{TIMSS2011}. This circumstance led to inquiry-based activities by junior and senior high school students with the cooperation of universities and research facilities. 
There is a lack of research by junior and senior high school students in the field of particle and atomic physics in particular. For example, out of 363 abstracts in Science Castle 2019, which is one of the largest conferences in Asia for junior and senior high school students, not one focused on atomic or particle research~\cite{science-castle}.
This is likely because atomic and particle physics research requires large and expensive equipment, such as large accelerators, which is difficult to prepare in schools.
Hence, to promote the research activities in the field of particle and atomic physics, it is important for accelerator facilities to provide opportunities for junior and senior high school students to engage in research of particle or atomic physics with their accelerators.

\par
In addition, research experience using an accelerator is also important from the perspective of radiation education. In Japan, radiation education has been included in the curriculum of junior high schools since the Curriculum Guidelines in 2008~\cite{curriculum}, and since the Fukushima nuclear accident in 2011, radiation education has become more important. It is difficult for students to learn and understand invisible radiation from textbooks alone, and continuous learning experiences through research activities are required~\cite{10.1093/jrr/rry025}.
Given this situation, providing opportunities for junior and senior high school students to conduct research at accelerator facilities, which have come to play a variety of roles using radiation, including radiation therapy as well as physics research, is important.
\par
Beamline for Schools~\cite{Gutowski_2018, Broomfield_2018, Biesot_2016}, organized by the European Organization for Nuclear Research (CERN) and Deutsches Elektronen-Synchrotron (DESY), is an interesting example of the use of a large accelerator. This is a competition for high school students from all over the world where they propose a scientific experiment that they want to perform with a particle accelerator. The two teams that prepare the best proposals win a trip to a particle accelerator facility to perform their experiments. In 2019, there were 178 proposals from 49 countries.
However, globally, there are no other examples that provide such opportunities.
\par
We made the first attempt to organize a similar workshop---``Accel Kitchen'' ---for junior and senior high school students by using the cyclotron accelerator in Cyclotron and Radioisotope Center (CYRIC). 
In this workshop, participants take part in the ongoing beam experiment with researchers for two days.
We held the first workshop in December 2018 and the second workshop in July 2019 that included the experiment of francium (Fr) production and laser trapping of Fr atoms. In June 2020, we experimented with the time-of-flight measurement of the proton beam provided for the radiation resistance test.
Due to the Covid-19, participants joined this proton experiment to manipulate the beam remotely.
In this study, we focus on the first two workshops.

\section{\label{sec:outline}Outline of workshops}
The first two workshops aimed to optimize the laser trap for the Fr atom (atomic number 87) that is rarely found on Earth. We needed to produce Fr atoms in each experiment with the accelerator due to the limited half-life time of $^{223}$Fr, which is about 22 minutes. 
There are previous works that produce and laser trap Fr for spectroscopy like SUNY Stony Brook~\cite{PhysRevLett.76.3522}, LNL Legnaro~\cite{ATUTOV2004421}, and TRIUMF~\cite{Tandecki_2013}, but CYRIC is the only facility that works on the development of laser trapping of Fr in Japan.
Therefore, this research project is a valuable opportunity for junior and senior high school students to experience the production of heavy atoms by nuclear reaction in a large accelerator.

\par
\begin{figure}[htbp]
  \begin{center}
    \includegraphics[width=\hsize,keepaspectratio]{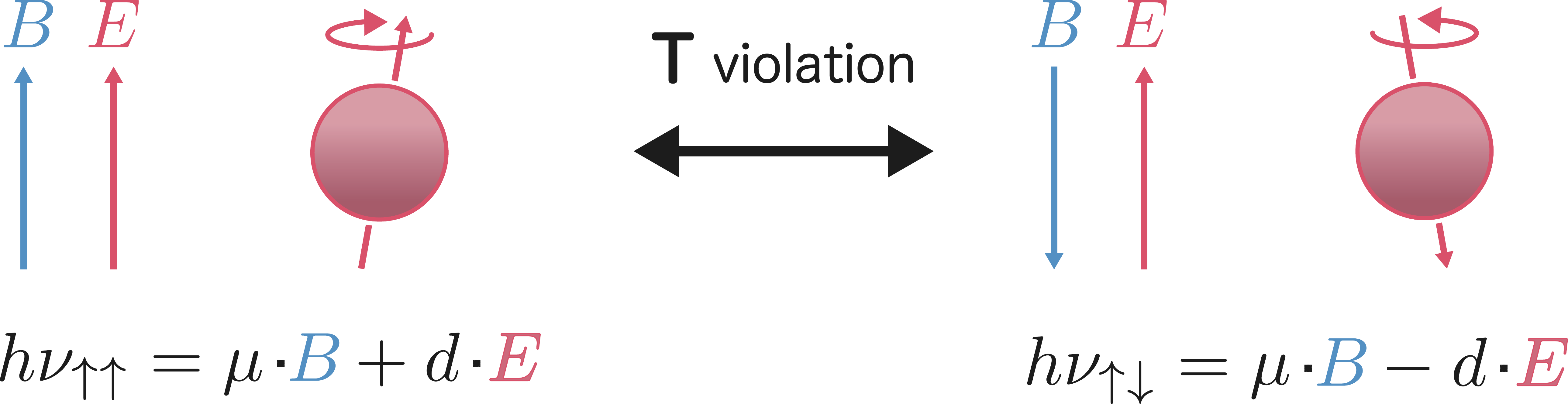}
    \caption{In order to extract the EDM, the Larmor precession of the spin with the magnetic moment ($\mu$) in the presence of parallel and antiparallel magnetic($B$) and electric fields($E$) is measured. The EDM is obtained from the energy difference of the two cases($h\nu_{\uparrow\uparrow}-h\nu_{\downarrow\uparrow}$).}
    \label{fredm-purpose.eps}
  \end{center}
\end{figure}

The main purpose of Fr production in CYRIC is to search for the electron electric dipole moment (e-EDM). The experimental searches for the electron permanent e-EDM is a sensitive probe to understand the violation of the fundamental symmetries such as CP and T, where, C~-~charge conjugation, P~-~parity, and T~-~time-reversal symmetry, respectively. The origin of the CP violation and the dark matter can be investigated with the e-EDM studies. In the Standard Model (SM) of particle physics, the predicted value of e-EDM, $|d_{e}| \approx 10^{-40}\ e \rm{cm}$, appears from the CP-violating components of the CKM matrix, and is too small to be observed in the currently ongoing EDM experiments~\cite{doi:10.1142/S0217751X12300153}. This means that the new physics can be implied if a finite value of the e-EDM can be measured with the current experimental sensitivity~\cite{POSPELOV2005119}. From a theoretical estimation, the Fr atom expresses a large enhancement factor (the ratio of atomic EDM to e-EDM) as $\sim 900$~\cite{PhysRevA.88.042507,Shitara2021}. Thus, Fr is one of the good candidates to search e-EDM.
An atomic EDM can be induced by the presence of an e-EDM, so an e-EDM can be obtained by measuring the atomic EDM of Fr.
The atomic EDM of Fr is extracted by measuring the Larmor precession of the spin with the magnetic moment ($\mu$) in the presence of parallel and antiparallel magnetic ($B$) and electric fields ($E$) (Fig.~\ref{fredm-purpose.eps}).
\par

\par
\begin{figure}[htbp]
  \begin{center}
  \includegraphics[width=\hsize]{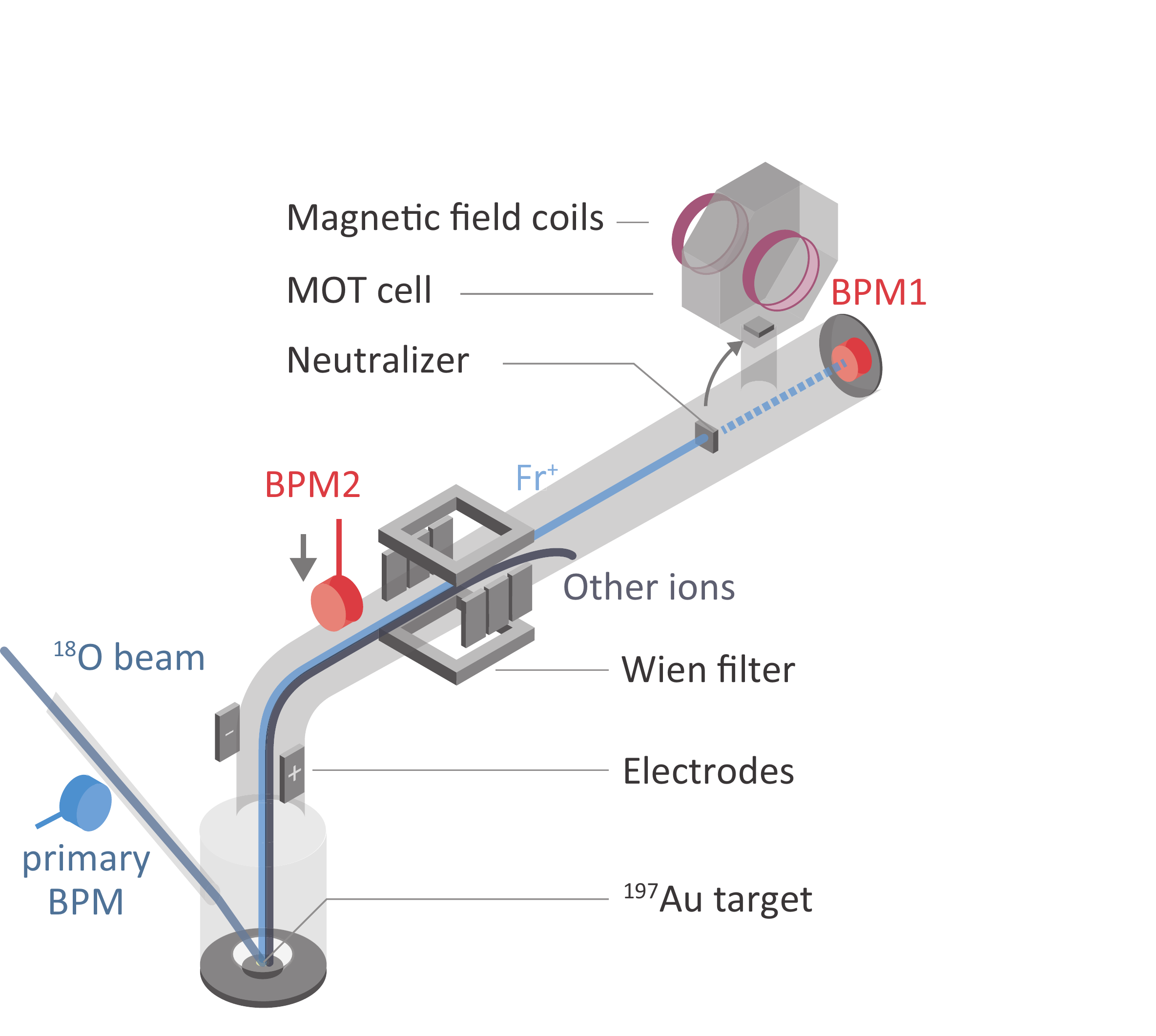}
  \caption{Schematic view of $\rm{Fr}^{+}$ beamline at CYRIC. $\rm{Fr}^{+}$ produced by the fusion-evaporation reaction at the gold target passes through the Wien mass filter and is transported to the yttrium (Y) target of dimensions 10~mm $\times$ 10~mm. Two beam profile monitors (BPM) consist of a microchannel plate (MCP), a phosphor screen, and  a charge-coupled device (CCD) camera are installed to observe the ion-beam profile~\cite{TANAKA2021165803}. BPM1 is installed at the end of the beamline and BPM2 is placed after the $90^{\circ}$ bending. The MOT cell is oriented $90^{\circ}$ (upward) towards the beamline. The BPM for monitoring the primary $^{18}\rm{O}$ beam is installed before the gold target.}
  \label{CYRIC-beamline.eps}
  \end{center}
  \end{figure}

The schematic of the setup developed at CYRIC for the production and trapping of Fr is shown in Figure~\ref{CYRIC-beamline.eps}. $^{209-215}\rm{Fr}$ isotopes are synthesized by the fusion-evaporation reaction between the oxygen ($^{18}\rm{O}$) beam from the Azimuthally Varying Field (AVF) cyclotron and gold ($^{197}\rm{Au}$) target~\cite{published_papers/21027008}.The Fr isotopes produced inside the gold target are simultaneously ionized at the surface of the gold and are extracted to approximately 12 m away from the production target to the trapping area. The flux of Fr is about $10^{6} /\mathrm{s}$ with a $^{18}\rm{O}$ beam current of $\approx$ 1~$\mu$A~\cite{Kawamura2015}. The $\rm{Fr}^{+}$ beam is irradiated and accumulated on an yttrium neutralizer (Y).  Thermal Fr atoms are released by heating the Y foil to about $10^{3}$~K~\cite{Kawamura2015}, and then collected in a magneto-optical trap (MOT)~\cite{Harada_2016}. These laser-cooled Fr atoms are used for the measurement of e-EDM~\cite{JPSConf.Proc.2.010112}. 

\par
In 2019 $\sim$ 2020, we succeeded in neutralizing the Fr on an yttrium neutralizer (Y) target, and the laser trapping of Fr was the next step.
Therefore, we planned ``Accel Kitchen’’ to improve the number of Fr provided to the trap region and try the laser trapping of Fr with the participants.
\subsection{System}
To support the participants’ activities, seven research collaborators and undergraduate students (2 freshmen and 1 sophomore for the first workshop; 4 freshmen and 2 sophomores for the second workshop) were part of the Fr experiment.
Undergraduate students took part in multiple offline experiments before the workshop.
In offline experiments, we used a rubidium ion gun instead of the Fr beam to practice the beam transportation and $^{241}$Am source to test the measurement of alpha-particle energy.
\subsection{Recruiting}
\begin{figure}[htbp]
  \begin{center}
    \includegraphics[width=\hsize,keepaspectratio,bb=0 0 1975 1481]{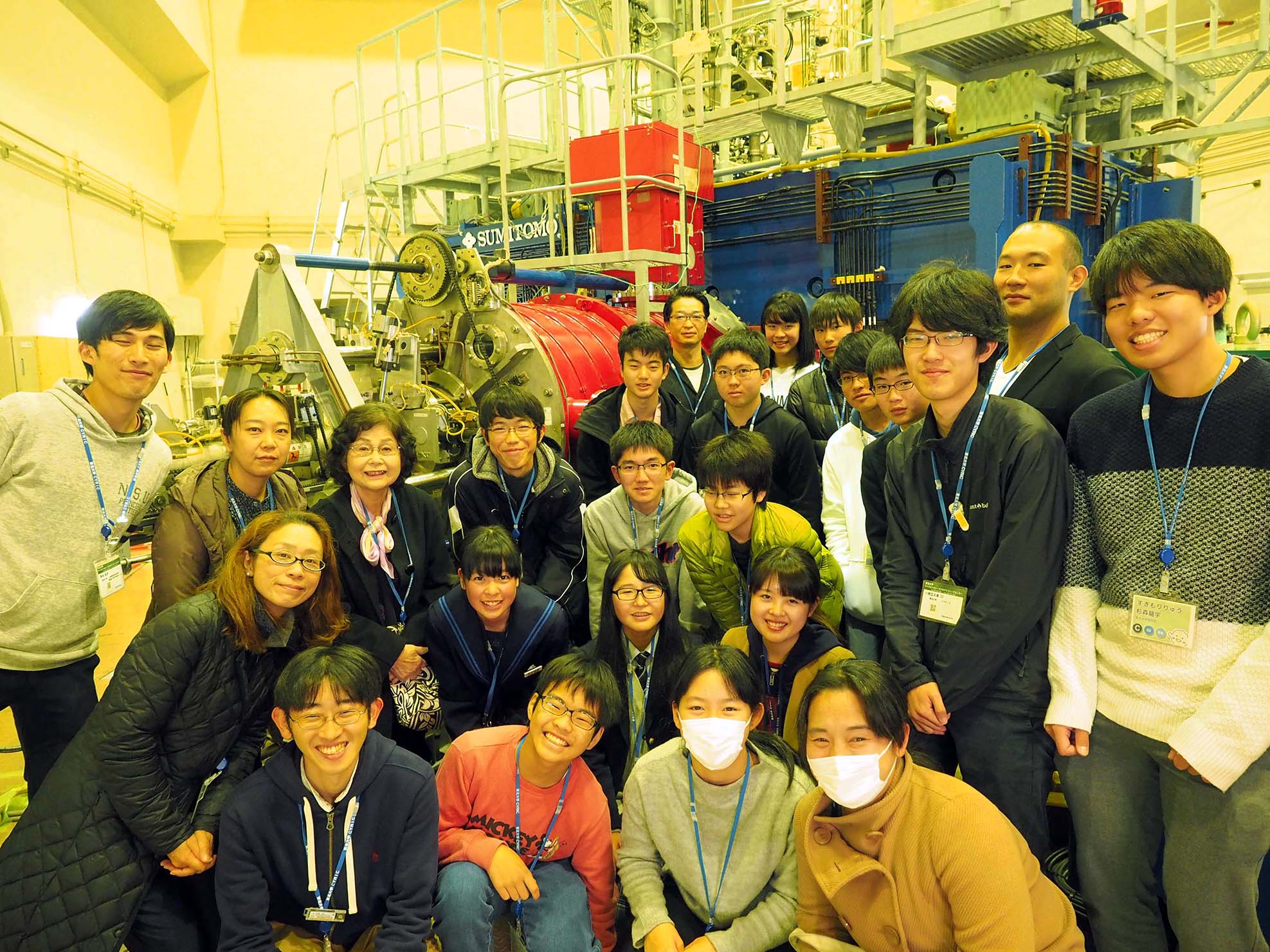}
    \caption{A group photo from the first workshop}
    \label{accel-kitchen-1st.jpg}
  \end{center}
  \end{figure}

\begin{figure}[htbp]
    \begin{center}
      \includegraphics[width=\hsize,keepaspectratio,bb=0 0 2268 1512]{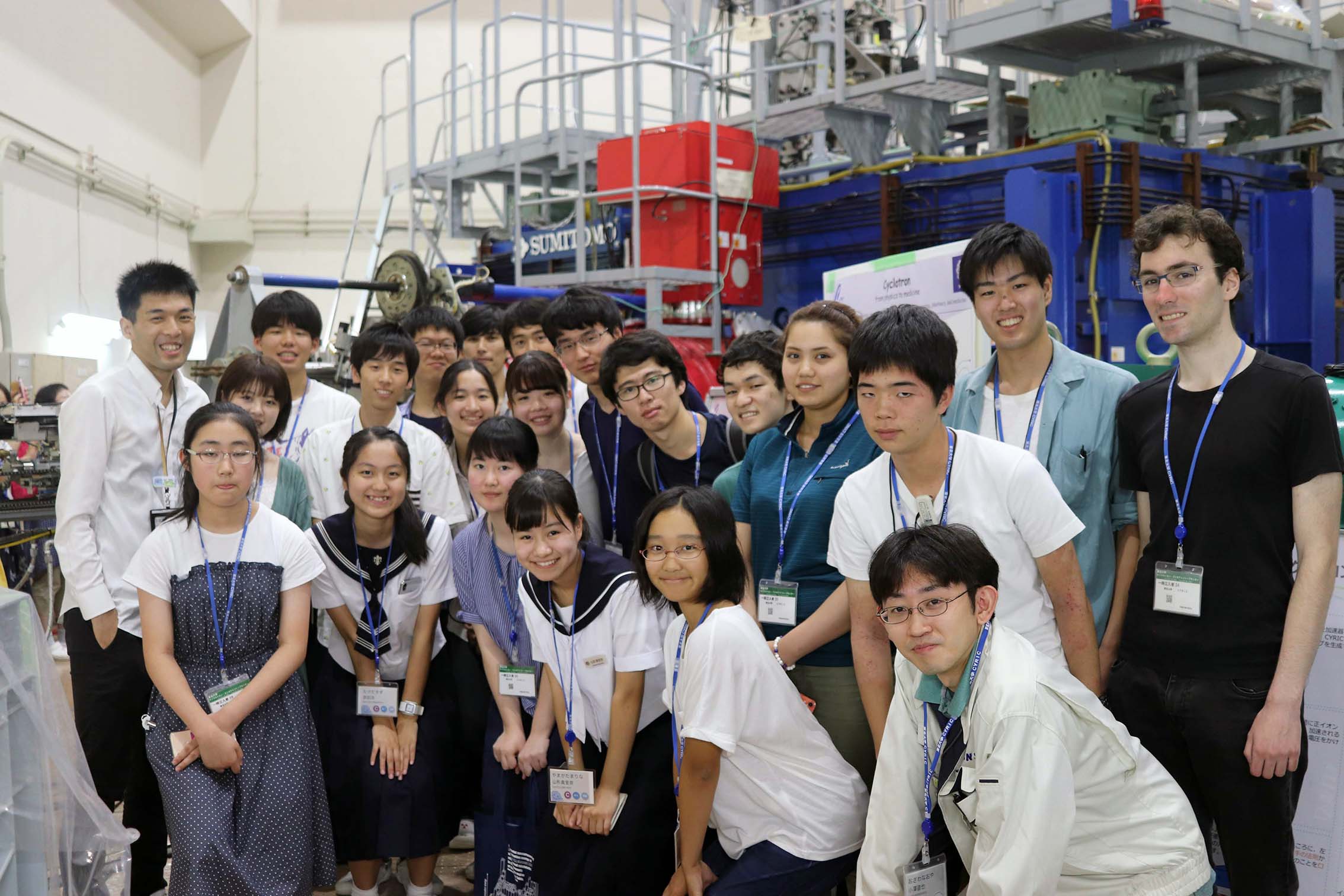}
      \caption{A group photo from the second workshop}
      \label{accel-kitchen-2nd.jpg}
    \end{center}
    \end{figure}

The first ``Accel Kitchen’’ was funded by Hirameki Tokimeki Science by Japan Society for the Promotion of Science. Hirameki Tokimeki Science aims to channel financial support awarded in the form of Grants-in-Aid for Scientific Research (KAKENHI) back into the society. Twenty-four junior and senior high school students applied for the first workshop, of which 12 students were selected through a screening process (Fig.~\ref{accel-kitchen-1st.jpg}).
The second workshop was supported by CYRIC. We gathered junior and senior high school students through web media and local radio.
Fifty-two junior and senior high school students applied for the second workshop, of which 12 students were selected (Fig.~\ref{accel-kitchen-2nd.jpg}).

\subsection{Content}
\begin{figure}[htbp]
  \begin{center}
    \includegraphics[width=\hsize,keepaspectratio]{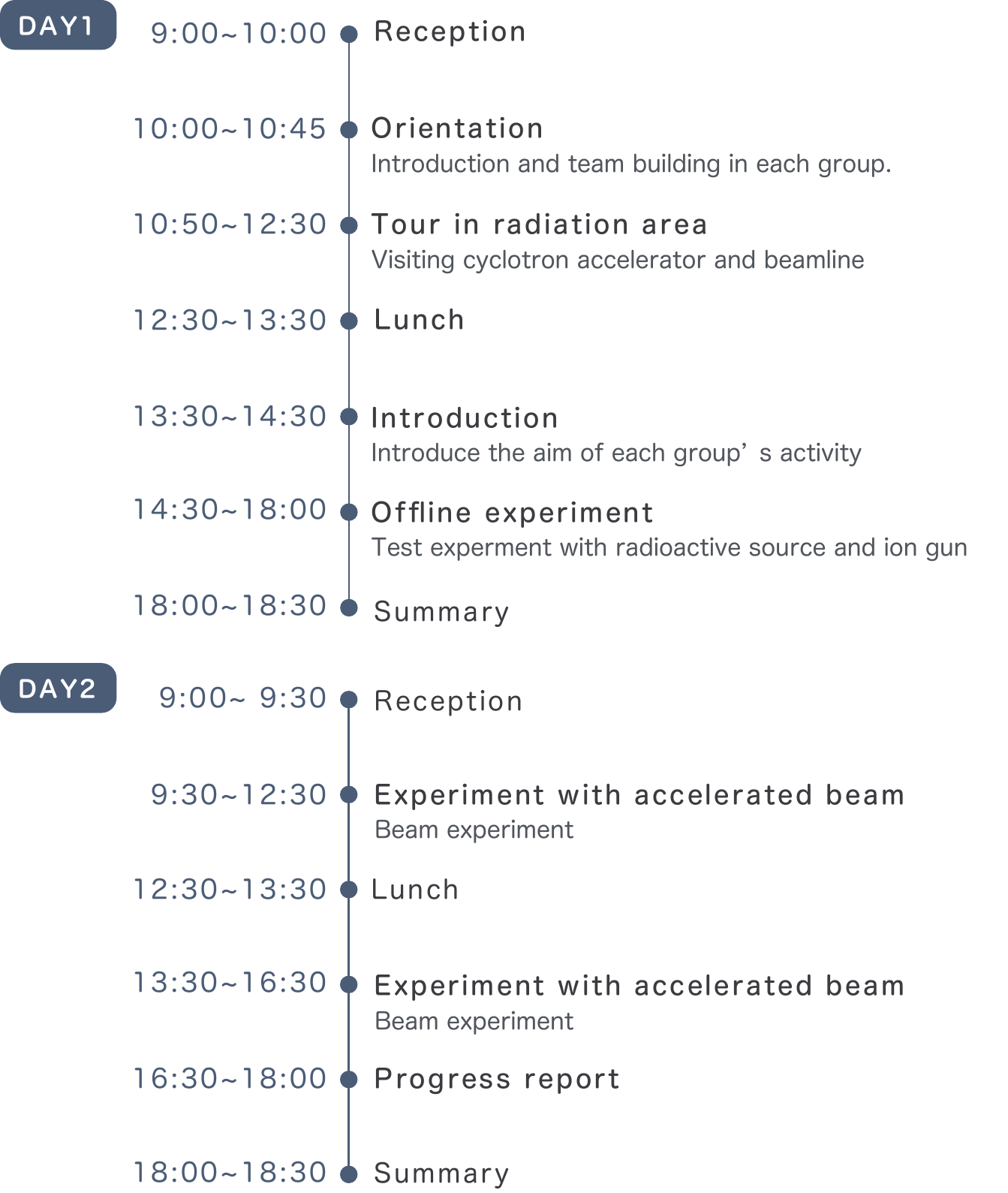}
    \caption{Schedule of the first workshop. Day~1: Visiting the experimental area and offline test experiments. Day~2: Experiment with the accelerated beam.}
    \label{schedule.eps}
  \end{center}
  \end{figure}

\begin{figure}[htbp]
\begin{center}
  \includegraphics[width=\hsize,keepaspectratio]{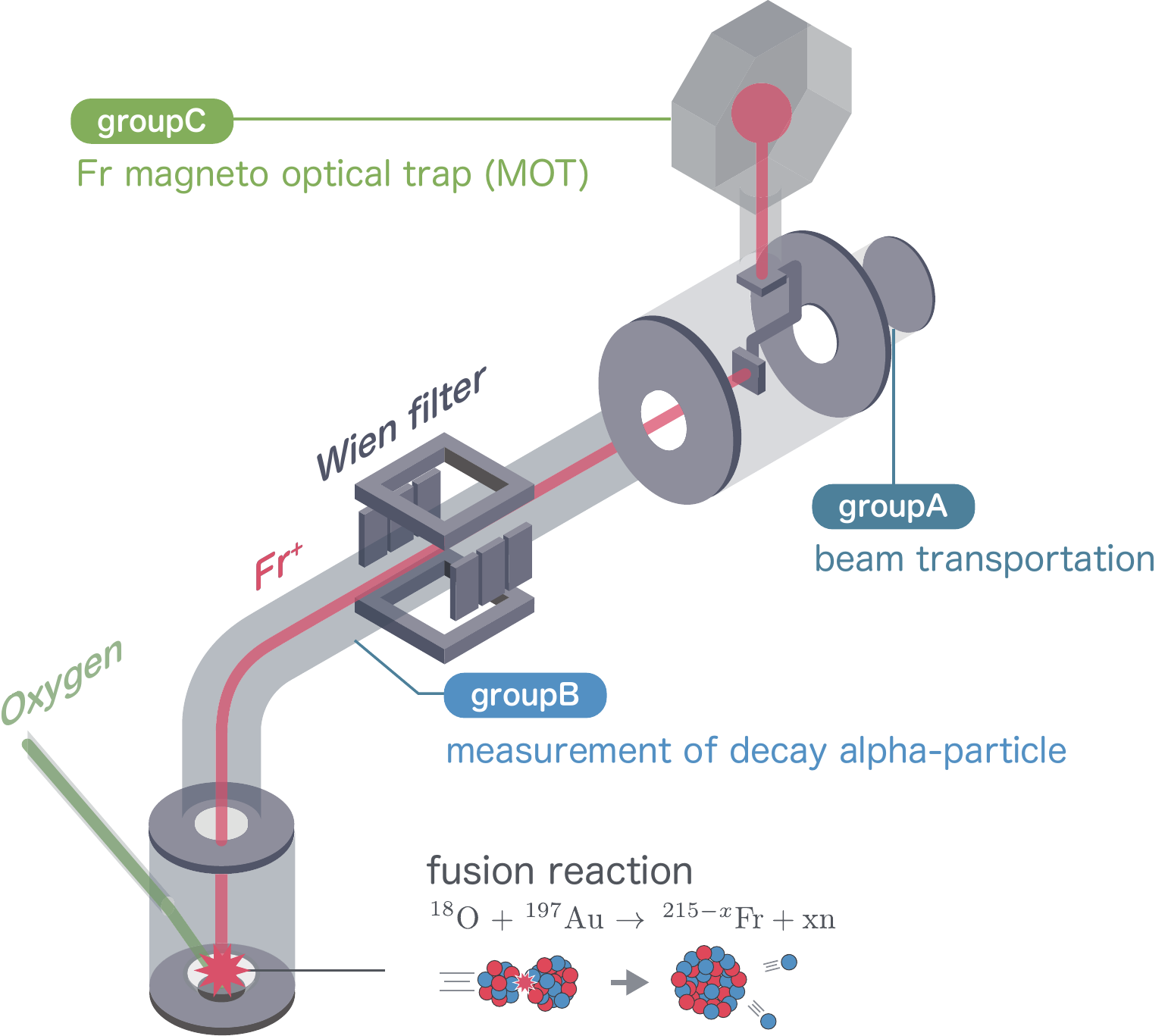}
  \caption{Overview of the research activities of each group. Group A delivered the Fr$^{+}$ beam to the end of the beamline and removed other ions with the Wien filter. Group B measured the number of each isotope of Fr that reached the end of the beamline by using a Solid-state detector (SSD). Group C optimized the condition of the magnetic field of the MOT for Fr atoms.}
  \label{experimental-purpose.eps}
\end{center}
\end{figure}

Participants were separated into three groups (Fig.~\ref{experimental-purpose.eps}). There were four junior and senior high school students, researchers, and undergraduate students in each group.
This was a two-day event (Fig.~\ref{schedule.eps}). Day~1 consisted of visiting the experimental area and conducting offline test experiments. Day~2 consisted of experiments with the accelerated beam.
Since participants were not registered as radiation workers, they could not enter the radiation area after beam acceleration.

\subsection{Beam transportation (Group A)}
\begin{figure}[htbp]
\begin{center}
  \includegraphics[width=\hsize,keepaspectratio]{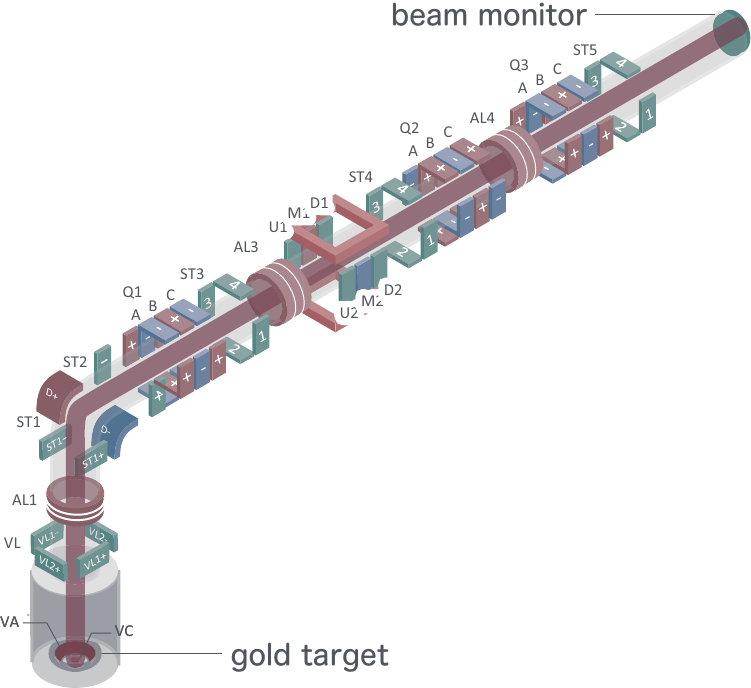}
  \caption{Fr beamline consists of steering electrodes(ST1-5), quadrupole electrodes for focusing(Q1-3), Einzel lens(AL1-4), and the Wien filter, which remove other ions except Fr$^{+}$.}
  \label{procedure2018-groupA.eps}
\end{center}
\end{figure}

\begin{figure}[htbp]
\begin{center}
  \includegraphics[width=\hsize,keepaspectratio]{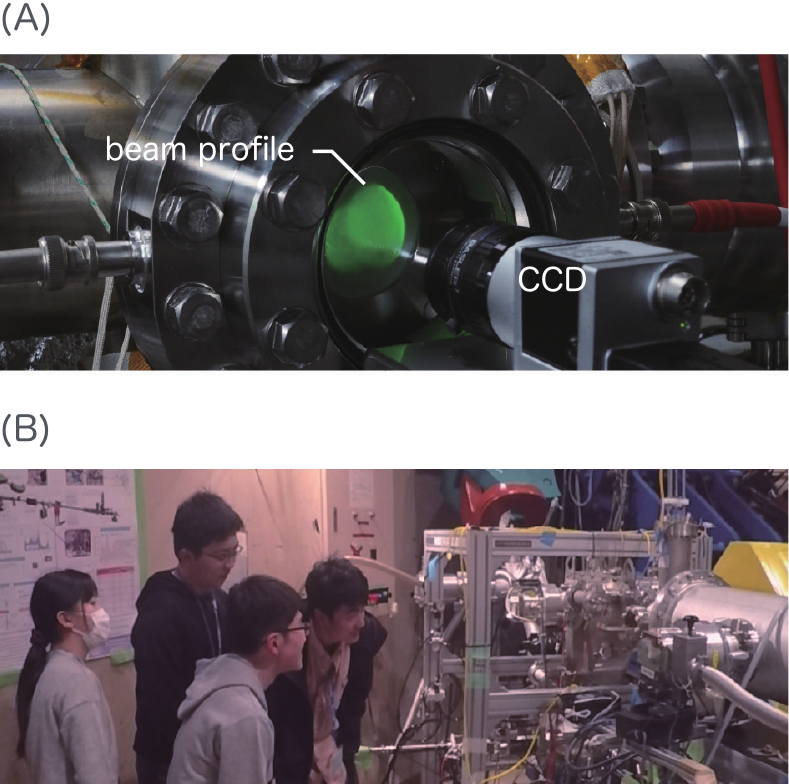}
  \caption{Group A activity. (A)The green glow of the phosphor screen can be observed when the Fr beam irradiates on it. This beam profile is recorded by the CCD camera. (B)Participants observed the beam profile from the rubidium ion gun.}
  \label{photo2018-groupA.eps}
\end{center}
\end{figure}

As Fig.~\ref{procedure2018-groupA.eps} shows, Group A transported the Fr beam to the BPM (Fig.~\ref{photo2018-groupA.eps}) located at the end of the beamline, 12 m away from the Fr production region with electrodes. 
\par
The Wien filter consists of perpendicular electric and magnetic field ($E$ and $B$), and only in case the Coulomb force ($qE$ where $q$ is the charge of ion) and Lorentz force ($qvB$ where $v$ is the velocity and $m$ is the mass of the ion) are equivalent, the ions can pass through it.
The condition can be expressed as
\begin{equation}
  qEd = q (v \times  B),
\end{equation}
where $d$ is a gap of electrodes of 60 mm.
Since ions are accelerated by the voltage of the gold target ($E_{ext}$), the kinetic energy of the ion is 
\begin{equation}
  \frac{1}{2} mv^2 = qE_{ext},
\end{equation}
so that the charge to mass ratio of ions that is allowed to pass through is 
\begin{equation}
  \frac{q}{m}= \frac{(Ed)^2}{2E_{ext}B^2}. \label{eq-qmass-ratio}
\end{equation}
\par
\begin{figure}[htbp]
\begin{center}
  \includegraphics[width=\hsize,keepaspectratio]{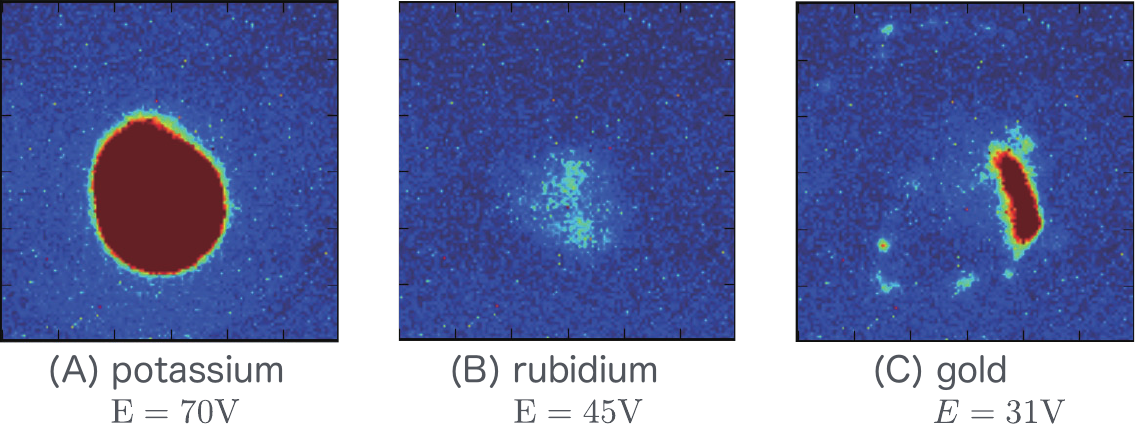}
  \caption{Beam profiles of (A)potassium ($E = 70$ V), (B)rubidium ($E = 45$ V) and (C)gold ($E = 31$ V) selected by Wien filter}
  \label{result2018-groupA.eps}
\end{center}
\end{figure}

In this workshop, we set the voltage of the gold target as $E_{ext} = 1$~kV and the magnetic field of Wien filter as $B = 0.03$~T and scan the electric field $E$ to identify the ions produced on the gold target.
Figure.~\ref{result2018-groupA.eps} shows an observed beam profile on BPM by applying three different voltages to the electrode $E$. Only the ions which have the charge to mass ratio calculated by Eq.~\ref{eq-qmass-ratio} can be observed.
Participants identified (A)potassium ($E = 70$~V), (B)rubidium ($E = 45$~V) and (C)gold ($E = 31$~V) with the Wien filter.
Potassium originated in the impurities of constructional elements of the apparatus.
Rubidium adhered to the target at the offline measurement. Rubidium and gold evaporated due to heating of gold caused by the oxygen beam collision with the target.

\subsection{Measurement of decay alpha-particle (Group B)}
\begin{figure}[htbp]
\begin{center}
  \includegraphics[width=\hsize,keepaspectratio]{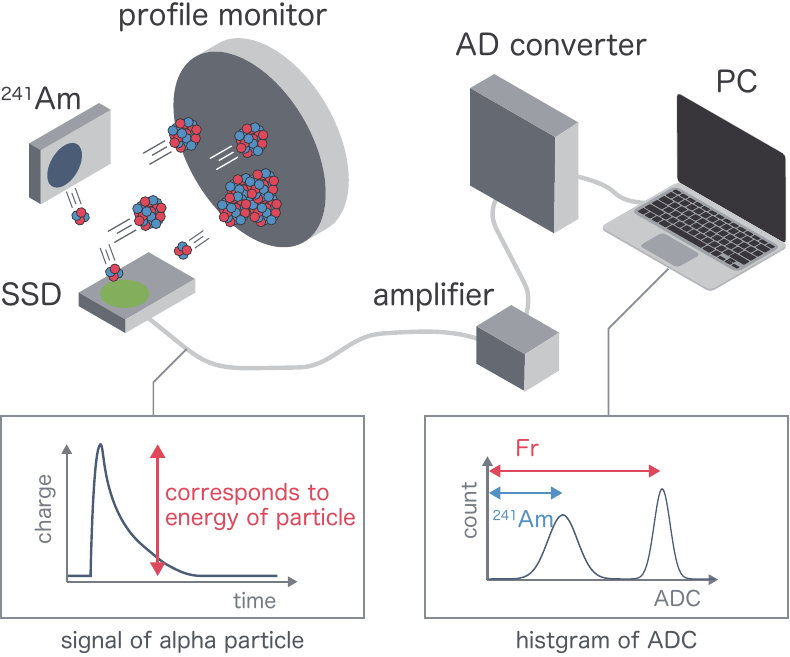}
  \caption{The schematic diagram of the way to identify the alpha-decay isotopes. Alpha-decay isotopes are accumulated from the irradiation of the Fr beam and radioactive source ($^{241}$Am), which is for the energy calibration of SSD. SSD converts the energy of alpha particles into charge and alpha-decay isotopes. The charge by each signal was converted to ADC value by the analog-digital converter and the histogram of the ADC was plotted to identify the energy of alpha particles in the isotopes.}
  \label{procedure2018-groupB.eps}
\end{center}
\end{figure}

\begin{figure}[htbp]
\begin{center}
  \includegraphics[width=\hsize,keepaspectratio]{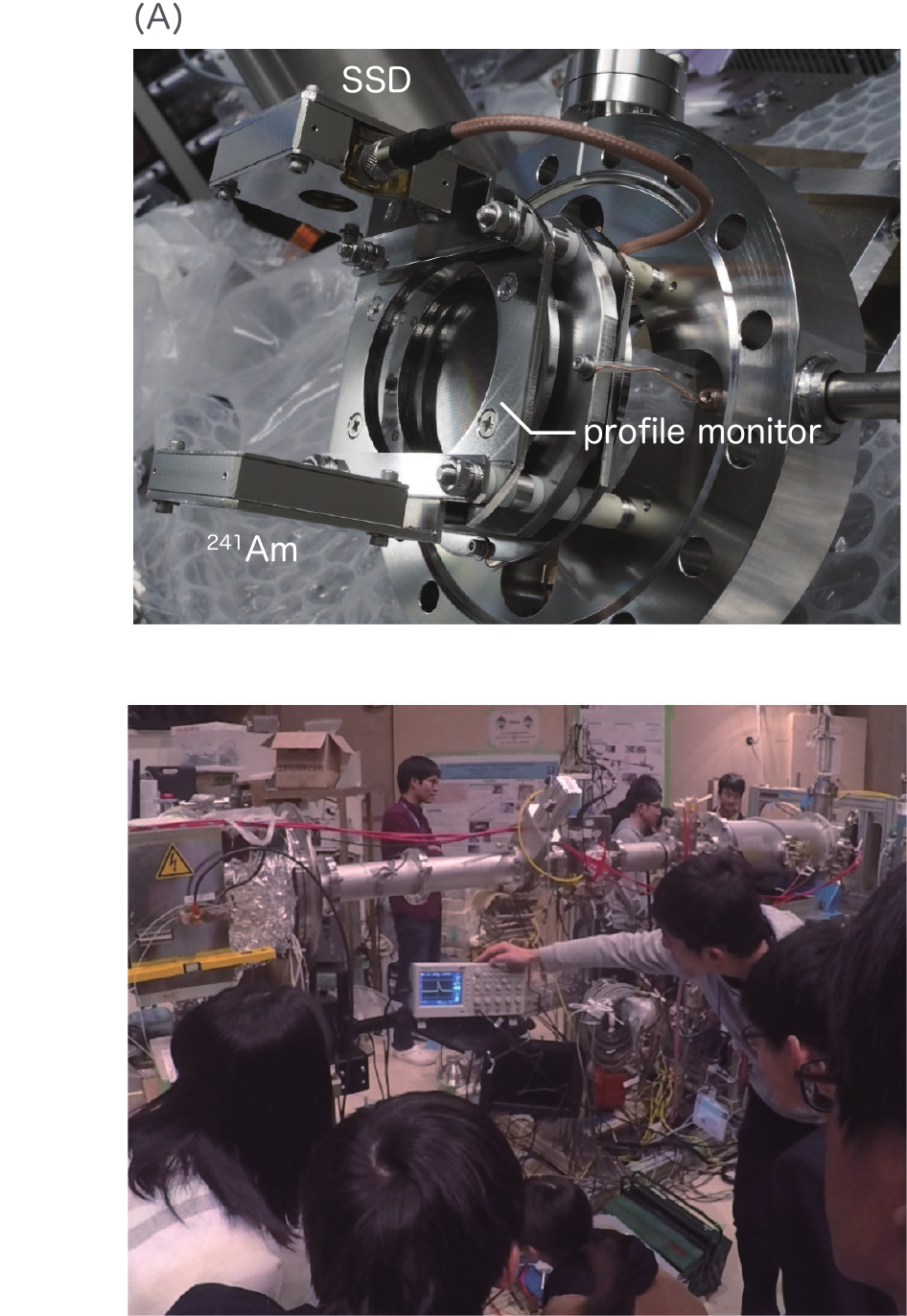}
  \caption{SSD and radioactive source ($^{241}$Am) are installed in front of the BPM to detect the alpha particles emitted from its surface (B)Participants observed the signal from SSD with an oscilloscope.}
  \label{photo2018-groupB.eps}
\end{center}
\end{figure}

\begin{figure}[htbp]
\begin{center}
  \includegraphics[width=\hsize,keepaspectratio]{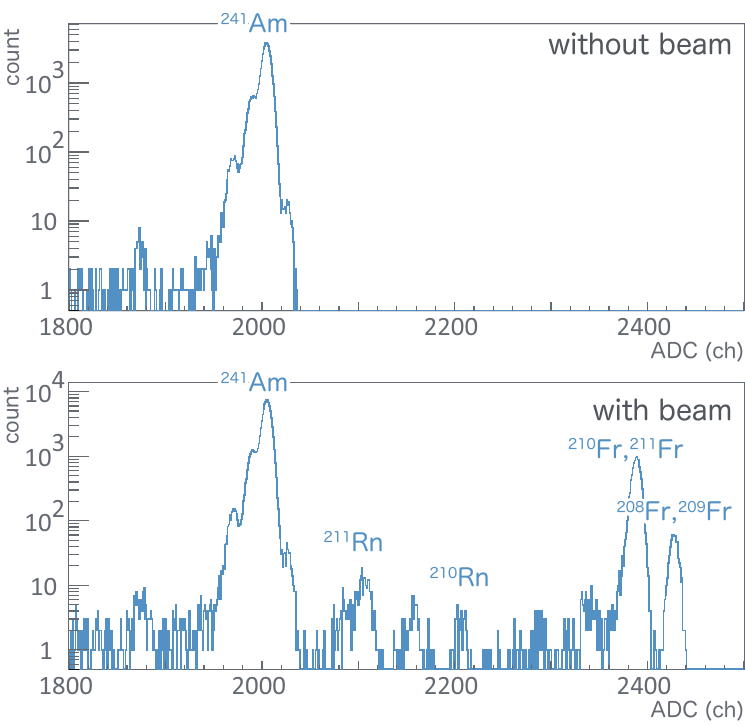}
  \caption{The energy spectrum was measured by Group B without a beam and with a beam. Alpha decay from a radioactive source($^{241}$Am) is observed without a beam. With irradiation of beam, they observed alpha decay from Fr isotopes ($^{208-211}$Fr) and daughter isotope such as $^{210-211}$Rn.}
  \label{result2018-groupB.eps}
\end{center}
\end{figure}

Group B measured the energy spectrum of the decay alpha emitted from the surface of the BPM with the silicon-based SSD (Fig.~\ref{procedure2018-groupB.eps}). There are various alpha-decay isotopes produced by the nuclear fusion reaction of an oxygen ($^{18}\rm{O}$) beam and a gold ($^{197}\rm{Au}$) target, and we can identify isotopes by measuring the kinetic energy of alpha particle from the decays, e.g., 6.5~MeV from $^{210}\rm{Fr}$~\cite{SHAMSUZZOHABASUNIA2014561}.
To estimate the beam flux of Fr$^{+}$, Group B measured the number of decay alpha particles by using the SSD which has an energy that corresponds to that from Fr (Fig.~\ref{photo2018-groupB.eps}).
Figure.~\ref{result2018-groupB.eps} is an energy spectrum measured by Group B without a beam and with a beam.
They identified each peak according to the NuDat2~\cite{doi:10.1063/1.1945075}.

There are peaks from $^{241}\rm{Am}$, which are from the test source in both spectrums.
After the beam irradiation, they observed peaks from $^{208}\rm{Fr}$, $^{209}\rm{Fr}$ and $^{210}\rm{Fr}$, $^{211}\rm{Fr}$ and $^{211}\rm{Rn}$,$^{210}\rm{Rn}$, which are daughter isotopes of Fr.

During the beam irradiation, Fr atoms are constantly supplied to BPM from an Fr beam of intensity $f$. At the same time, the Fr atoms on the surface of BPM decay with the mean lifetime $\tau$. The number $N$ of $\rm{Fr}^{+}$ on the BPM is expressed as
\begin{equation}
  \frac{\rm{d}N}{\rm{d}t}  =  f - \frac{1}{\tau},
\end{equation}
Of these, few are short-lived, such as, $^{208}\rm{Fr}$ ($\tau = 85.3$~s) and $^{209}\rm{Fr}$ ($\tau = 72.9$~s), while others are long-lived, such as $^{210}\rm{Fr}$ ($\tau = 275.3$~s) and $^{211}\rm{Fr}$ ($\tau = 268.3$~s)~\cite{SHAMSUZZOHABASUNIA2014561,MARTIN20071583,CHEN2015373,SINGH2013661}.
The number of Fr particles during beam irradiation is expressed as
\begin{eqnarray}
  N_{\rm{on}} & = & \tau f (1-\exp{\frac{t-t_{0}}{\tau}}) +N(t_{0}) \exp{(-\frac{t-t_{0}}{\tau})},  \label{eq:N_on}  \\
  N_{\rm{off}} & = & \left[ \tau f\exp{(\frac{t_{1}}{\tau})} + N(t_{0}) \exp{(\frac{t_{0}}{\tau})} \right] \exp{(-\frac{t}{\tau})}.\label{eq:N_off} 
\end{eqnarray}
 
\begin{figure}[htbp]
  \begin{center}
    \includegraphics[width=\hsize]{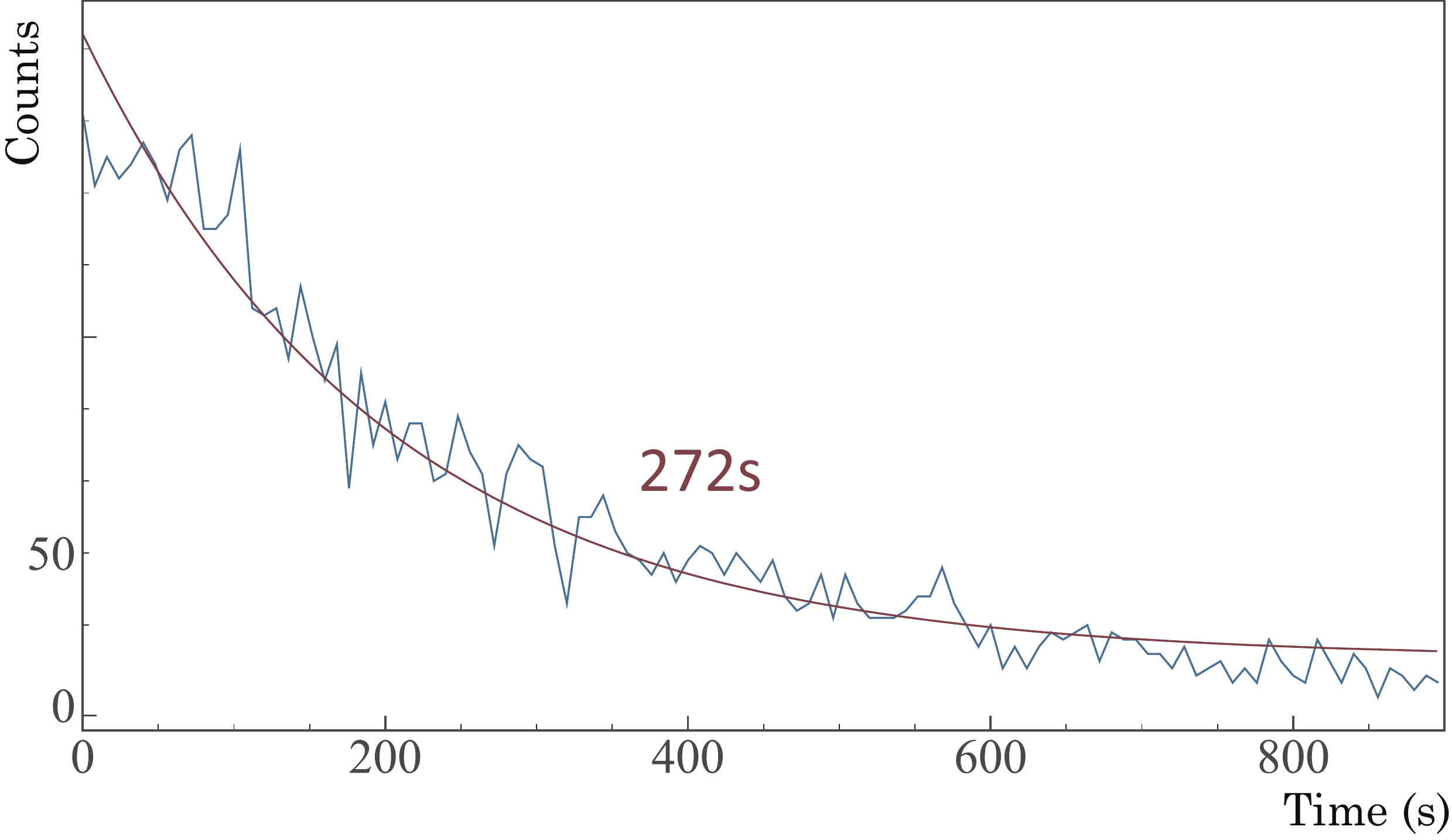}
    \caption{
      The time structure of the SSD counts that the energy of decay alpha particle corresponds to $^{210}\rm{Fr}$ or $^{211}\rm{Fr}$ after stopping the ion beam.
      The lifetimes of $^{210}\rm{Fr}$ and $^{211}\rm{Fr}$ were estimated as $\tau = 272 \pm 13$~s by fitting them with an exponential function (red line).}
    \label{run24-ssd-decay.eps}
  \end{center}
\end{figure}

The Fr beam flux, $f$, and the lifetime of Fr, $\tau$, can be estimated by fitting it to Eq.~\ref{eq:N_on} and Eq.~\ref{eq:N_off}, respectively. 
Further, the time structure of the SSD counts that the energy of decay alpha particle corresponds to $^{210}\rm{Fr}$ or $^{211}\rm{Fr}$ after stopping the ion beam is shown in Fig.~\ref{run24-ssd-decay.eps}. The lifetime of Fr was extracted by fitting Eq.~\ref{eq:N_off} to the measured result. The estimated sum lifetime of $^{210}\rm{Fr}$ and $^{211}\rm{Fr}$ from the SSD counts was $\tau = 272\pm 13$~s.

\subsection{Laser trapping of Fr (Group C)}
\begin{figure}[htbp]
\begin{center}
  \includegraphics[width=0.7\hsize,keepaspectratio]{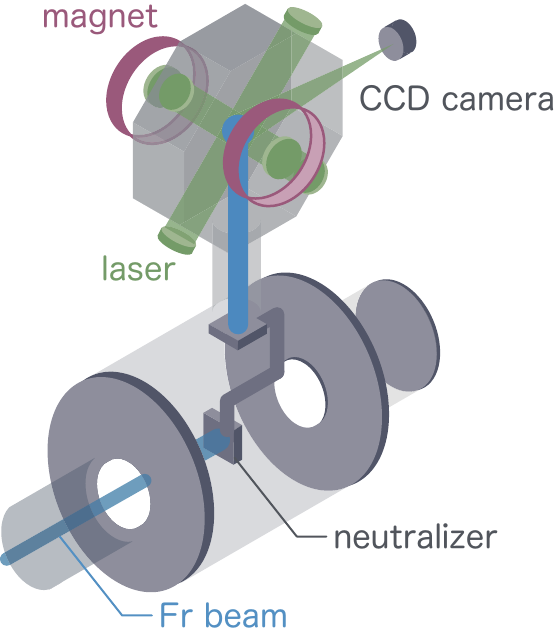}
  \caption{The schematic view of the laser-trapping system. Upon heating the yttrium foil accumulate the Fr atoms to approximately $10^{3}$ K, thermal and neutralized Fr atoms are released, which are then collected in a MOT. MOT is a combination of six circularly-polarized, red-detuned optical beams and a gradient of  magnetic field.
  They could observe fluorescence from trapped atoms by using a CCD camera just after applying the magnetic field, and the number of trapped atoms is estimated from the fluorescence strength.}
  \label{procedure2018-groupC.eps}
\end{center}
\end{figure}

Group C tried the MOT of Fr atoms transported by groups A and B (Fig.~\ref{procedure2018-groupC.eps}).
However, they could not observe the trapped atoms due to the lack of Fr atoms delivered to the trap system. Therefore, they switched their aim from trapping Fr atoms to optimizing the applied magnetic field by using rubidium atoms, which are provided by the dispenser.

MOT is an apparatus that uses laser cooling and a spatially varying magnetic field to create a trap that can produce samples of cold, trapped, and neutral atoms.

\begin{figure}[htbp]
\begin{center}
  \includegraphics[width=\hsize,keepaspectratio]{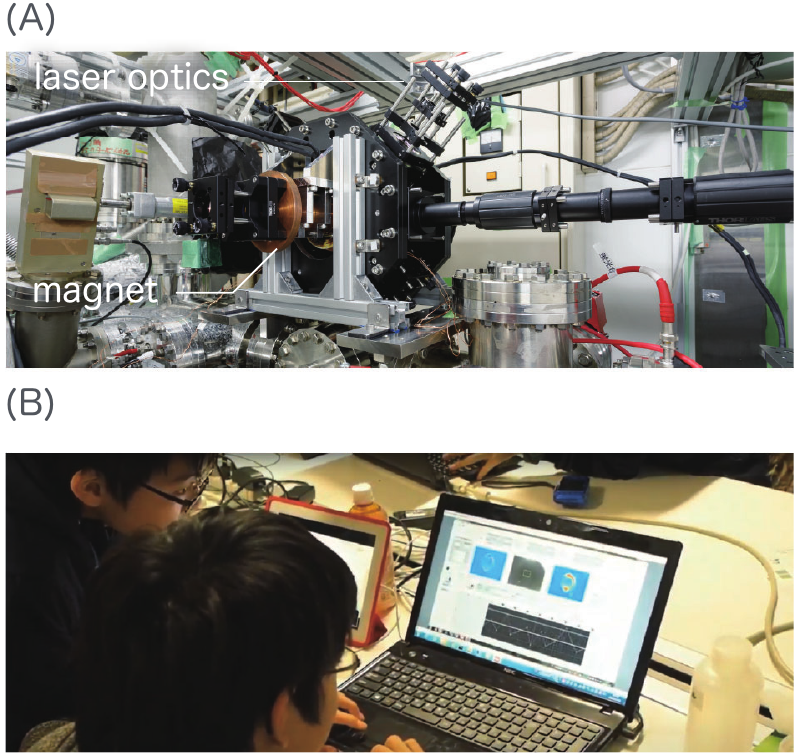}
  \caption{(A) The photograph of MOT consists of the magnet, the six optical laser beams, and the CCD camera. (B)Participants analyze the observed images from the CCD camera.}
  \label{photo2018-groupC.eps}
\end{center}
\end{figure}

\begin{figure}[htbp]
\begin{center}
  \includegraphics[width=\hsize,keepaspectratio]{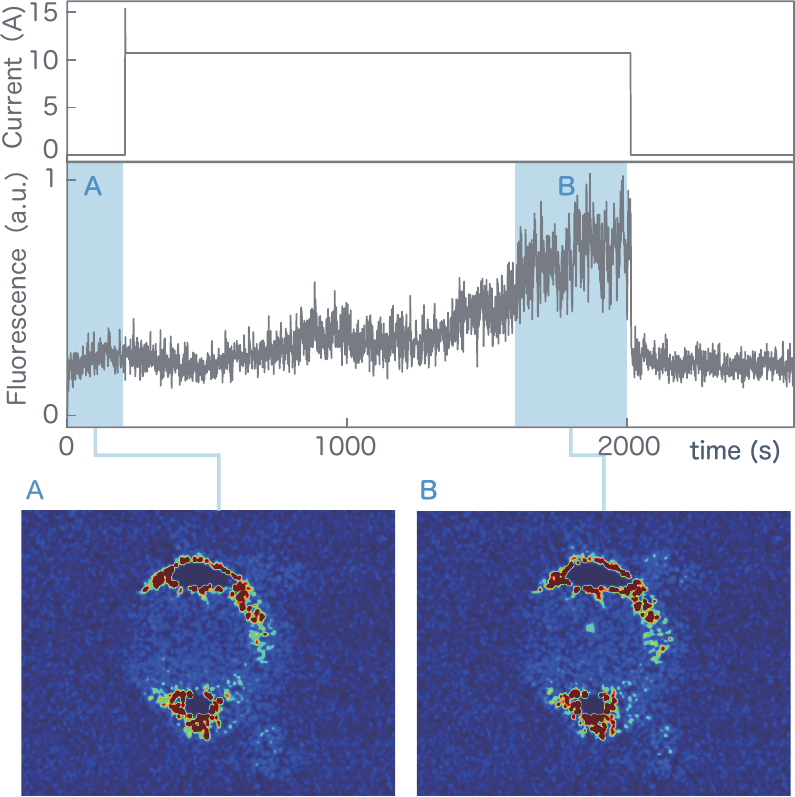}
  \caption{The fluorescence is increased by accumulating trapped rubidium atoms in MOT after applying a gradient magnetic field. Compared to the image from CCD, before applying the field (A), trapped atoms are observed for about 30 minutes after applying the field (B).}
  \label{result2018-groupC.eps}
\end{center}
\end{figure}

\begin{figure}[htbp]
  \begin{center}
    \includegraphics[width=\hsize,keepaspectratio]{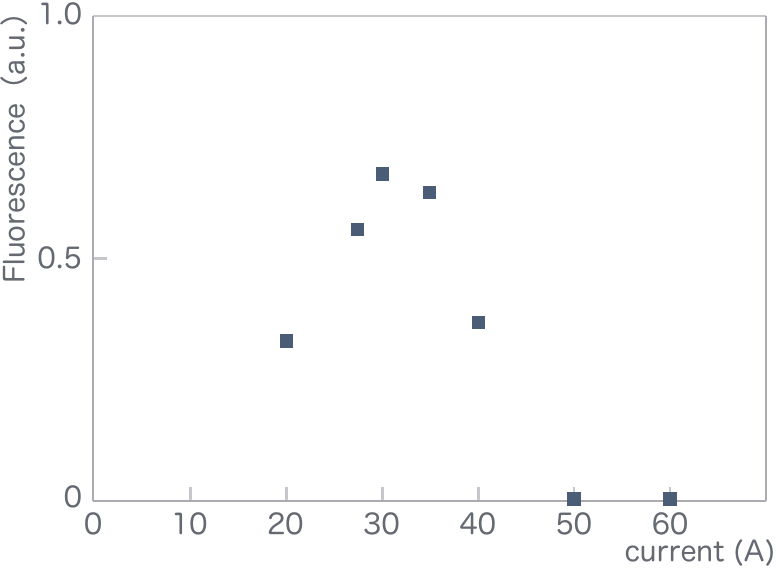}
    \caption{Variation of the intensity of fluorescence obtained by CCD as a function of the current application to the magnet. The optimum current to maximize the number of trapped atoms is about 30 A.}
    \label{result2018-groupC2.eps}
  \end{center}
  \end{figure}

MOT is a combination of six, circularly-polarized, red-detuned optical laser beams, and a gradient of a magnetic field~\cite{PhysRevLett.59.2631}.
They can observe fluorescence from trapped atoms by using a CCD camera just after applying the magnetic field. The number of trapped atoms is estimated from the fluorescence strength (Fig.~\ref{photo2018-groupC.eps}).
If the strength of the magnetic field is increased, the trapping potential gets deeper, but the trapping region gets narrower.
Group C measured the dependence of the magnetic field on a number of trapped atoms to find an optimal value of the magnetic field (Fig.~\ref{result2018-groupC.eps} and Fig.~\ref{result2018-groupC2.eps}).


  \subsection{Progress report}
\begin{figure}[htbp]
    \begin{center}
      \includegraphics[width=\hsize,keepaspectratio,bb=0 0 2345 1563]{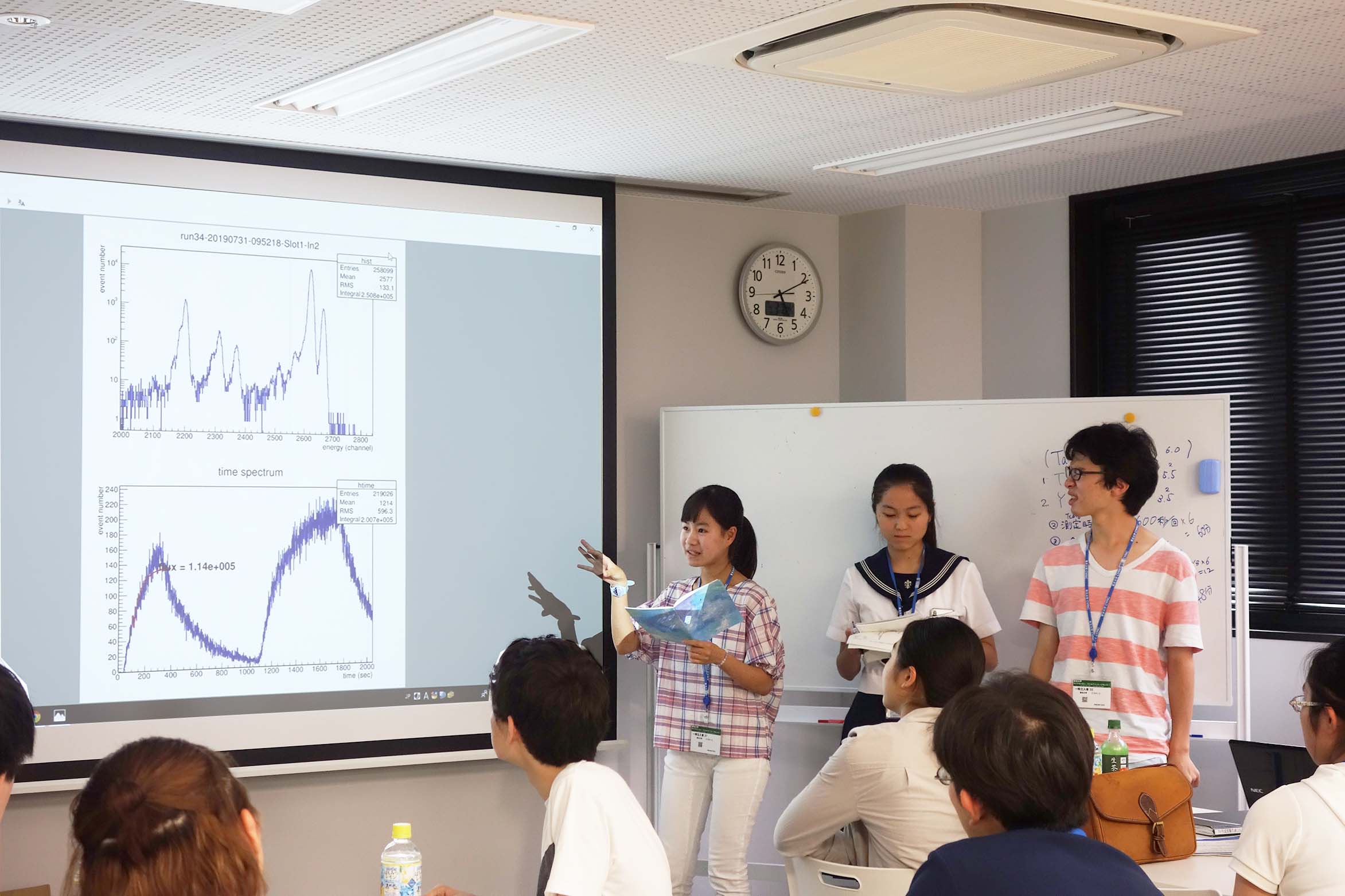}
      \caption{Photograph of the progress report}
      \label{summary-report.jpg}
    \end{center}
    \end{figure}

  Each group reported their result at the end of the workshop (Fig.~\ref{summary-report.jpg}) and discussed it with the researchers. To allow maximum research time, participants were forbidden from preparing any slides for presentation and were only allowed to use plots for describing their results.

\section{\label{sec:conclusion}Conclusion}
We held ``Accel Kitchen’’ in 2018 and 2019 at Tohoku University for junior and senior high school students to provide an opportunity to participate in an ongoing particle or atomic experiment by using an accelerator. 
Twelve participants in each workshop aimed to laser trap Fr atoms with researchers and undergraduate students over two days. The activity motivated the participants’ interests for accelerator physics and radiation. 
\par
After the workshop, we gathered feedback from participants. The following was written by one of participants who later entered the Department of Chemistry, Faculty of Science Tohoku University. 
\begin{quote}
  Before the workshop, I only knew about X-rays in terms of radiation but I learned that there are various ways of using them, even producing heavy atoms that do not exist on Earth. Now I am considering stepping into the field of chemistry using radiation isotopes.
\end{quote}

\par
The following was written by a participant who later entered the Department of Physics, Faculty of Science, Tohoku University to study particle physics.
\begin{quote}
  I was only aware that radiation was invisible and dangerous, just like a virus, before the workshop. But I learned that radiation is familiar in our society.
  I aspire to enter Tohoku University through this experience to learn about this good environment to engage in particle-physics research with its large accelerator.
\end{quote}
\par
The following was expressed by a participant who later entered the Department of Radiological Technology, School of Health Sciences, Faculty of Medicine to study nuclear medicine.
\begin{quote}
  My view on career paths was influenced by this experience. Before the workshop, I vaguely want to enter a medical field, but afterward, I was interested in particle physics and radiation because of this event. My aspiration for a career became clear: to study nuclear medicine.
\end{quote}
  
In 2021, four prior participants from these two workshops, as undergraduate students who had graduated from high school, joined the members of ``Accel Kitchen’’ to support the next junior and senior high school students and participated in the experiment in CYRIC. This workshop thus gave them the opportunity to have an impact on the field of radiation and accelerators for junior and senior high school students and influence their career paths.
\par
``Accel Kitchen’’ was introduced as a representative attempt by Hirameki Tokimeki Science in 2019~\cite{hiratoki}.
Results from ``Accel Kitchen’’ contributed to the Fr experiment in CYRIC; after two workshops, the first trapping of Fr was achieved in CYRIC by using the optimum magnetic field by Group C~\cite{doi:10.1063/5.0037134}.
In addition, Group A evaluated the performance of the BPMs, and this diagnostic system was included in a published paper ~\cite{TANAKA2021165803}.

\section{\label{sec:acknowledgement}Acknowledgement}
This research was supported by Hirameki Tokimeki Science by JSPS and JSPS KAKENHI, Grant Numbers 26220705 and 19H05601.

\section{\label{sec:ethical-statement}Ethical Statement}
We obtained consent to publish from all student's parents or guardians prior to the workshops.

\bibliographystyle{ptephy}
\bibliography{phys-educ-accel-kitchen}

\end{document}